\documentclass{article}
\usepackage{spconf,amsmath,graphicx}
\usepackage{booktabs}
\usepackage{color,amsmath,url,times,tabularx,bbm,amssymb,multirow}
\usepackage{bm}
\usepackage[numbers,sort&compress]{natbib}

\title{Improving Text-Audio Retrieval by Text-aware Attention Pooling and Prior Matrix Revised Loss}

\name{Yifei Xin,
      Dongchao Yang,
      Yuexian Zou$^{*}$\thanks{This paper was partially supported by NSFC (No: 62176008) and Shenzhen Science \& Technology Research Program (No:GXWD20201231165807007-20200814115301001).}\thanks{$^{*}$ Yuexian Zou is the corresponding author.}}
\address{School of ECE, Peking University, Shenzhen, China}
\begin{document}
%
\maketitle
\begin{abstract}
In text-audio retrieval (TAR) tasks, due to the heterogeneity of contents between text and audio, the semantic information contained in the text is only similar to certain frames within the audio. Yet, existing works aggregate the entire audio without considering the text, such as mean-pooling over the frames, which is likely to encode misleading audio information not described in the given text. In this paper, we present a text-aware attention pooling (TAP) module for TAR, which is essentially a scaled dot product attention for a text to attend to its most semantically similar frames. Furthermore, previous methods only conduct the softmax for every single-side retrieval, ignoring the potential cross-retrieval information. By exploring the intrinsic prior of each text-audio pair, we introduce a prior matrix revised (PMR) loss to filter the hard case with high (or low) text-to-audio but low (or high) audio-to-text similarity scores, thus achieving the dual optimal match. Experiments show that our TAP significantly outperforms various text-agnostic pooling functions. Moreover, our PMR loss also shows stable performance gains on multiple datasets.
\end{abstract}
\section{Introduction}
\label{sec:intro}
Given a caption or an audio clip as a query, the text-audio retrieval (TAR) task aims at retrieving a paired item from a set of candidates in another modality. To compute the similarity between the two modalities, a common technique is to embed a text and an audio clip into a shared latent space and then adopt a distance metric like the cosine similarity to measure the relevance of the text and audio. 

However, there is a significant disparity between both modalities that makes such a direct interaction challenging, that is, the heterogeneity of contents across different modalities \cite{li2021otcmr,zhang2021hcmsl,gorti2022x}. Specifically, the semantic information contained in the text is typically similar to sub-segments of an audio clip. In this case, common text-agnostic aggregation schemes that pool entire audio frames, such as mean-pooling, might encode redundant or even distracting acoustic information that is not described in the given text. Moreover, depending on the input text, the most semantically similar frames would vary, so there could be multiple equally valid texts that match a specific audio clip. Therefore, we would expect the same audio to be retrieved for any of these queries and a retrieval model to prioritize the audio sub-segments that are most pertinent to the provided text. 

Besides, previous TAR methods only perform the softmax operation along a single dimension for each retrieval pair \cite{xie2022dcase,lai2022resnet}, which ignores the potential cross-retrieval information and harms the retrieval performance. To solve this, we introduce the dual optimal match hypothesis based on the discovered phenomenon from previous extensive experiments \cite{lou2022audio,koepke2022audio,cheng2021improving} that when a text-to-audio or audio-to-text pair reaches the optimal match (single-side match), the symmetric audio-to-text or text-to-audio scores should be the highest. With this hypothesis, a prior matrix revised (PMR) loss is introduced to revise the similarity matrix between the text and audio. Specifically, we first introduce a prior probability matrix calculated in the cross direction to adjust the original similarity score. Then, by conducting the dot product of the prior probability matrix and original scaling similarity matrix, we can filter the case with a high text-to-audio (or audio-to-text) similarity score but a low audio-to-text (or text-to-audio) similarity score, thus achieving the dual optimal match and leading to a more convincing result.

The major contributions of this paper are summarized as follows: 
\begin{itemize} 
\item We present a text-aware attention pooling (TAP) module that allows a model to reason about the most relevant audio frames to a provided text while suppressing frames not described in the given text.
\item We introduce a prior matrix revised (PMR) loss to revise the similarity matrix between the text and audio, which imposes a direct constraint to filter those single-side match pairs and highlights more convincing results with the dual optimal match. 
\item Experiments show that our TAP significantly outperforms text-agnostic audio pooling functions. Furthermore, our PMR loss also shows stable performance gains on multiple datasets.
\end{itemize}
\begin{figure}[t]
  \centering
  \includegraphics[width=1.0\linewidth]{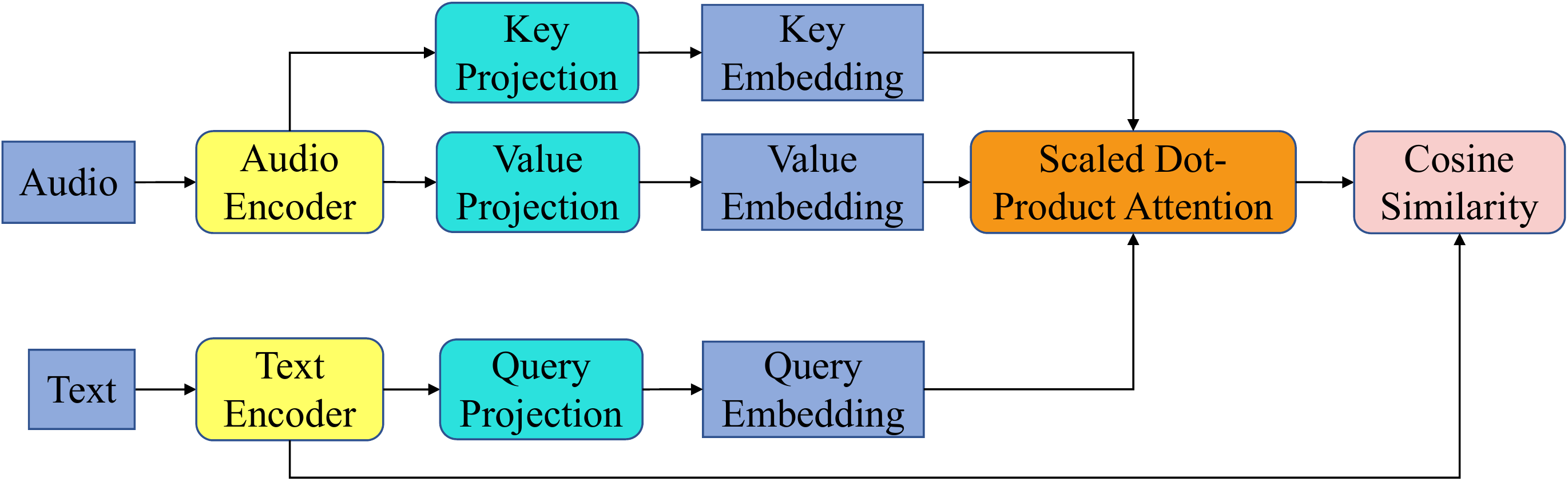}
  \caption{Overview of our text-aware attention pooling module.}
  \label{fig:adapt-pic}
  \vspace*{-\baselineskip}
\end{figure}
\section{Problem Formulation}
\label{sec:method}
We define two text-audio retrieval (TAR) tasks, where the text-to-audio retrieval is denoted as $t2a$ and the audio-to-text retrieval is denoted as $a2t$. In $t2a$, we are provided with a query text and an audio set. The target is to rank all audio clips according to their similarities with the query text. Similarly, in $a2t$, we are provided with a query audio clip and a text set. The aim is to retrieve matched texts based on their relevance with the query audio.

The TAR models usually consist of a text encoder (e.g., BERT-styled models that achieve superior performance on various NLP tasks \cite{kenton2019bert}) and an audio encoder (e.g., pretrained audio tagging networks \cite{kong2020panns,chen2022hts,xin2022audio}), which project the text and audio into a shared embedding space, respectively. Specifically, given a text $t$ and an audio clip $a$ as input, the text encoder outputs the text embedding $c_t \in \mathbb{R}^{D}$, while the audio encoder is employed to generate the audio embedding $c_a \in \mathbb{R}^{T \times D}$, where $D$ is the size of the model’s channel dimension and $T$ is the number of audio frames. In order to embed our given text and audio into a shared space to compute the similarity score, an aggregation function $p(\cdot)$ (e.g., mean-pooling) is utilized to pool the frame-level feature $c_a$ into the clip-level latent embedding $z_a \in \mathbb{R}^{D}$: 
\begin{equation}
             z_a = p(c_a), \quad z_t = c_t.
\end{equation}
Therefore, the similarity of the text and audio can be measured by the cosine similarity of their embeddings:
\begin{equation}
             s(t,a) = \frac{z_t \cdot z_a}{\left\| z_t \right\| \left\| z_a \right\|}.
\end{equation}
Currently, the NT-Xent loss \cite{mei2022metric,chen2020simple} based on symmetrical cross-entropy is widely employed, which has been shown to consistently outperform the previous triplet-based losses \cite{xie2022dcase,lamorttake} on both $t2a$ and $a2t$ tasks. Therefore, we adopt it as the baseline loss function for our work. The NT-Xent loss is formulated as below:
\begin{equation}
\begin{aligned}
    L = -\frac{1}{B} \left(\sum_i^B {\rm log}\frac{{\rm exp}(s(t_i,a_i)/\tau)}{\sum_j^B {\rm exp}(s(t_i,a_j)/\tau)}+ \right.\ \\ \left. \sum_i^B {\rm log}\frac{{\rm exp}(s(t_i,a_i)/\tau)}{\sum_j^B {\rm exp}(s(t_j,a_i)/\tau)} \right),
\end{aligned}
\end{equation}
where $B$ is the batch size, $i$ and $j$ denote the sample index in a batch, and $\tau$ is a temperature hyper-parameter. The training objective is to maximize the similarity of the positive pair relative to all negative pairs within a mini-batch, and the ultimate loss is calculated in both directions.

\section{Proposed Methods}
\label{sec:proposed method}
\subsection{Text-aware Attention Pooling}
In existing TAR works, the aggregation function $p$ does not directly consider the input text and is merely a function of the audio frames such as max-pooling, mean-pooling schemes \cite{lou2022audio}. However, a text is most semantically related to the sub-segments of an audio clip. What's more, there could be multiple texts matching a specific audio clip, but the frames that are most semantically similar would vary. As such, text-agnostic aggregation functions would capture superfluous and distracting information not stated in the text, which impairs the TAR performance. Therefore, it is important to match a given text with its most semantically similar audio frames. 

To that end, we present a learnable text-aware attention pooling (TAP) module $\psi$ for TAR to perform the cross-modal reasoning on the audio frames that are most semantically related to a given text. The core mechanism is a scaled dot product attention \cite{gorti2022x,vaswani2017attention} between the text $t$ and the frames of an audio clip $a$. By conditioning $\psi$ on $t$, we can generate the audio aggregated embedding that learns to capture the most semantically similar audio frames as described in $t$. The resulting aggregated audio embedding is denoted as $z_{a|t}$, and our similarity function $s(t,a)$ is defined as:
\begin{equation}
             z_{a|t} = \psi(c_a|t),\quad s(t,a) = \frac{z_t \cdot z_{a|t}}{\left\| z_t \right\| \left\| z_{a|t} \right\|}.
\end{equation}
To elaborate, as shown in Fig. 1, we first project a text embedding $c_t \in \mathbb{R}^{D}$ output by the text encoder into a query $Q_t \in \mathbb{R}^{1 \times D_p}$ and an audio embedding $c_a \in \mathbb{R}^{T \times D}$ generated by the audio encoder into key $K_a \in \mathbb{R}^{T \times D_p}$ and value $V_a \in \mathbb{R}^{T \times D_p}$ matrices, where $D_p$ is the size of the projection dimension. The projections are defined as:
\begin{equation}
             Q_t = {\rm LN}(c_t^T)W_Q,
\end{equation}
\begin{equation}
             K_a = {\rm LN}(c_a)W_K,
\end{equation}
\begin{equation}
             V_a = {\rm LN}(c_a)W_V,
\end{equation}
where $LN$ represents a layer normalization layer \cite{ba2016layer} and $W_Q$, $W_K$ and $W_V$ are projection matrices in $\mathbb{R}^{D \times D_p}$. In order to flexibly learn the relevance between the given text and the audio frames, we then adapt the scaled dot product attention from the query-projected text embedding to the key-projected frame embedding. The dot product attention provides relevance weights from a text to each audio frame, which we adopt to aggregate the value-projected frame embedding: 
\begin{equation}
             {\rm Attention}(Q_t,K_a,V_a) = {\rm softmax}(\frac{Q_t K_a^T}{\sqrt{D_p}})V_a.
\end{equation}
Specifically, the query-projected text embedding is utilized to search frames with high relevance from the key-projected frame embedding. The value-projected embedding represents the audio’s context, from which we aggregate frames conditioned on the given text. To embed an audio clip into a shared space with a text, we project the aggregated audio feature from the attention module back into $\mathbb{R}^{D}$ by leveraging a weight $W_O \in \mathbb{R}^{D_p \times D}$ to obtain: 
\begin{equation}
             z_{a|t} = {\rm LN}({\rm Attention}(Q_t,K_a,V_a)W_O),
\end{equation}
where the resulting output $z_{a|t}$ is an aggregated audio embedding depending on the text $t$. By introducing the text-aware attention pooling, the model can concentrate on the most pertinent audio frames as described in a given text, thus effectively mitigating the negative impacts of the heterogeneity of contents between text and audio. Next, we will introduce a prior matrix revised loss to further revise the results for both $t2a$ and $a2t$ tasks.

\subsection{Prior Matrix Revised Loss}
Previous TAR loss functions \cite{mei2022metric,mei2022language,koh2022language} (e.g., the NTXent loss) only conduct the softmax for every single-side retrieval, which is just inferred with the similarity score for each row in the original similarity matrix, thus ignoring the potential cross-retrieval information. 

Based on the introduced dual optimal match hypothesis from previous extensive TAR experiments \cite{lou2022audio,koepke2022audio,mei2022metric} that when a $t2a$ (or $a2t$) pair reaches the single-side match, the symmetric $a2t$ (or $t2a$) score should also be the highest, a prior probability matrix is introduced to be calculated in the cross direction for $t2a$ and $a2t$ to fully exploit the cross-retrieval information. Specifically, the prior matrix is obtained by calculating the softmax score along each column of the original similarity matrix and then we multiply the prior matrix with the original similarity matrix to revise the similarity score, that is, for the $t2a$ (or $a2t$) task, we first calculate the softmax score of $a2t$ (or $t2a$), and then incorporate it into the loss calculation of $t2a$ (or $a2t$). Lastly, we conduct the softmax operation along each row of the revised similarity matrix to get the final probability result. Our prior matrix revised (PMR) loss is formulated as below:
\begin{equation}
\begin{aligned}
    L_{t2a} = -\frac{1}{B} \sum_i^B {\rm log}\frac{{\rm exp}(s(t_i,a_i) \cdot Pr_{i,i}^{t2a}/\tau)}{\sum_j^B {\rm exp}(\bm{s(t_i,a_j)} \cdot Pr_{i,j}^{t2a}/\tau)},
\end{aligned}
\end{equation}
\begin{equation}
\begin{aligned}
    L_{a2t} = -\frac{1}{B} \sum_i^B {\rm log}\frac{{\rm exp}(s(t_i,a_i) \cdot Pr_{i,i}^{a2t}/\tau)}{\sum_j^B {\rm exp}(\bm{s(t_j,a_i)} \cdot Pr_{j,i}^{a2t}/\tau)},
\end{aligned}
\end{equation}
\begin{equation}
\begin{aligned}
    L = L_{t2a} + L_{a2t},
\end{aligned}
\end{equation}
where $Pr^{t2a}$, $Pr^{a2t}$ denote the prior matrix for text-to-audio and audio-to-text tasks, respectively:
\begin{equation}
\begin{aligned}
    Pr_{i,j}^{t2a} = \frac{{\rm exp}(\omega \cdot s(t_i,a_i))}{\sum_j^B {\rm exp}(\omega \cdot \bm{s(t_j,a_i)})},
\end{aligned}
\end{equation}
\begin{equation}
\begin{aligned}
    Pr_{j,i}^{a2t} = \frac{{\rm exp}(\omega \cdot s(t_i,a_i))}{\sum_j^B {\rm exp}(\omega \cdot \bm{s(t_i,a_j)})},
\end{aligned}
\end{equation}
where $\omega$ represents a logit scaling parameter to smooth the gradients. In this way, $t2a$ and $a2t$ can coordinately revise each other's similarity scores, which provides prior knowledge of each other to filter the outliers and sharpen the more convincing points, thus achieving the dual optimal result. 
\begin{table}\small
  \caption{Performance comparison of our TAP and previous text-agnostic pooling methods.}
  \centering
  \label{tab:freq}
  \begin{tabular}{c|cc|cc}
    \toprule
    \multirow{2}{*}{Methods} & \multicolumn{2}{c|}{Text-to-Audio} & \multicolumn{2}{c}{Audio-to-Text} \\
    & \textbf{R@1} & \textbf{R@10} & \textbf{R@1} & \textbf{R@10}\\
    \midrule
    \multicolumn{5}{c}{\textbf{AudioCaps}}\\
    \midrule
    Mean & 32.6$\pm$0.5 & 81.0$\pm$0.4 & 37.9$\pm$0.8 & 82.4$\pm$0.8\\
    MeanMax & 33.9$\pm$0.4 & 82.6$\pm$0.3 & 39.4$\pm$1.0 & 83.9$\pm$0.6\\
    NetRVLAD & 34.5$\pm$0.4 & 83.4$\pm$0.7 & 40.1$\pm$1.0 & 84.3$\pm$0.6\\
    \textbf{TAP} & \textbf{36.1$\pm$0.2} & \textbf{85.2$\pm$0.4} & \textbf{41.3$\pm$0.5} & \textbf{86.1$\pm$0.3}\\
    CNN14+Mean & 29.8$\pm$0.5 & 78.5$\pm$0.4 & 40.3$\pm$0.7 & 81.3$\pm$0.3\\   
    \textbf{CNN14+TAP} & \textbf{33.1$\pm$0.4} & \textbf{81.3$\pm$0.8} & \textbf{42.8$\pm$0.4} & \textbf{84.1$\pm$0.2}\\
    \midrule
    \multicolumn{5}{c}{\textbf{Clotho}}\\
    \midrule
    Mean & 12.6$\pm$0.3 & 45.2$\pm$0.6 & 13.1$\pm$0.6 & 46.6$\pm$0.8\\
    MeanMax & 14.4$\pm$0.4 & 49.9$\pm$0.2 & 16.2$\pm$0.7 & 50.2$\pm$0.7\\
    NetRVLAD & 15.1$\pm$0.5 & 50.1$\pm$1.2 & 16.8$\pm$0.2 & 50.5$\pm$0.5\\
    \textbf{TAP} & \textbf{16.2$\pm$0.6} & \textbf{50.8$\pm$0.3} & \textbf{17.6$\pm$0.5} & \textbf{51.4$\pm$0.6}\\
    CNN14+Mean & 12.2$\pm$0.8 & 46.1$\pm$0.5 & 12.4$\pm$0.6 & 47.1$\pm$0.4\\   
    \textbf{CNN14+TAP} & \textbf{15.1$\pm$0.4} & \textbf{49.3$\pm$0.6} & \textbf{15.3$\pm$0.3} & \textbf{51.1$\pm$0.3}\\
  \bottomrule
\end{tabular}
\vspace*{-\baselineskip}
\end{table}
\section{Experiments}
\label{sec:exp}
\subsection{Dataset}
We evaluate our methods on two publicly available datasets: AudioCaps \cite{kim2019audiocaps} and Clotho \cite{drossos2020clotho} datasets. AudioCaps contains about 50K audio samples, which are all 10-second long. The training set consists of 49274 audio clips, each with one corresponding human-annotated caption. The validation and test sets contain 494 and 957 audio clips, each with five human-annotated captions. The Clotho v2 dataset contains 6974 audio samples between 15 and 30 seconds in length. Each audio sample is annotated with 5 sentences. The numbers of training, validation, and test samples are 3839, 1045, and 1045, respectively. 
\begin{table}
  \caption{Results of our PMR loss with previous methods.}
  \centering
  \label{tab:freq}
  \begin{tabular}{c|cc|cc}
    \toprule
    \multirow{2}{*}{Methods} & \multicolumn{2}{c|}{Text-to-Audio} & \multicolumn{2}{c}{Audio-to-Text} \\
    & \textbf{R@1} & \textbf{R@5} & \textbf{R@1} & \textbf{R@5}\\
    \midrule
    \multicolumn{5}{c}{\textbf{AudioCaps}}\\
    \midrule
    MeanMax+NTXent & 33.9 & 69.7 & 39.4 & 72.0\\
    \textbf{MeanMax+PMR} & \textbf{34.1} & \textbf{70.2} & \textbf{39.6} & \textbf{72.8}\\
    TAP+NTXent & 36.1 & 72.0 & 41.3 & 75.5\\
    \textbf{TAP+PMR} & \textbf{36.8} & \textbf{72.7} & \textbf{41.7} & \textbf{76.2}\\
    CNN14+TAP+NTXent & 33.1 & 68.6 & 42.8 & 72.7\\
    \textbf{CNN14+TAP+PMR} & \textbf{33.4} & \textbf{68.8} & \textbf{43.1} & \textbf{73.3}\\
    \midrule
    \multicolumn{5}{c}{\textbf{Clotho}}\\
    \midrule
    MeanMax+NTXent & 14.4 & 36.6 & 16.2 & 37.5\\
    \textbf{MeanMax+PMR} & \textbf{14.9} & \textbf{37.1} & \textbf{16.6} & \textbf{37.8}\\
    TAP+NTXent & 16.2 & 39.2 & 17.6 & 39.6\\
    \textbf{TAP+PMR} & \textbf{17.1} & \textbf{39.6} & \textbf{18.2} & \textbf{39.9}\\
    CNN14+TAP+NTXent & 15.1 & 36.7 & 15.3 & 36.5\\
    \textbf{CNN14+TAP+PMR} & \textbf{15.6} & \textbf{37.2} & \textbf{15.9} & \textbf{36.8}\\
  \bottomrule
\end{tabular}
\vspace*{-\baselineskip}
\end{table}
\subsection{Training Details and Evaluation metrics}
In our work, we follow the same pipeline in \cite{mei2022metric} to train our network. We adopt BERT \cite{kenton2019bert} as the text encoder, while employing the ResNet-38 in Pre-trained audio neural networks (PANNs) \cite{kong2020panns} as the audio encoder if not otherwise specified. We conduct experiments by fine-tuning the pre-trained models. The query, key and value projection dimension size is set as $D_p = 512$. Recall at rank k (R@k) is utilized as the evaluation metric, which is a popular cross-modal retrieval evaluation protocol. R@k measures the proportion of targets retrieved within the top-k ranked results, so a higher score means better performance. The results of R@1, R@5, and R@10 are reported.

\subsection{Experimental Results}
In this section, we first compare the performance of our TAP with various text-agnostic pooling functions on the AudioCaps and Clotho datasets. To demonstrate the superiority of our TAP, we compare it with previously popular aggregation strategies that achieve SOTA results, where Mean denotes the average pooling function, MeanMax \cite{mei2022metric} denotes we both use an average and max pooling layer to aggregate the frame-level feature, NetRVLAD \cite{lou2022audio,miech2017learnable} is a descriptor-based pooling method that enables back-propagation by adopting soft assignment to clusters, and TAP represents our text-aware attention pooling method. The experiments are repeated three times with different training seeds. 

As shown in Table 1, our TAP outperforms all other works that use text-agnostic pooling on all datasets and across all metrics, thereby highlighting the importance of our text-aware aggregation scheme that can learn to match a text with its most relevant frames while suppressing distracting information from irrelevant audio frames. In addition to adopting ResNet-38 as the backbone of our audio encoder, we also provide the results of using CNN14 \cite{kong2020panns} as our audio encoding model. It can be seen that our method also achieves performance boosts by a large margin, which strongly demonstrates the effectiveness and generalization of our TAP module. Notably, although the NetRVLAD aggregation method achieves relatively good results, it needs to manually select the number of clusters for different datasets and its tuning of the hyper-parameter is task and data specific. In contrast, our TAP can adaptively learn the optimal amount of information to extract for each text-audio pair, which thus removes the need to manually specify the hyper-parameter and can be more robust to different tasks and instances.

To evaluate our PMR loss, we compare it with the previous SOTA loss for TAR: NTXent, which has be shown in \cite{mei2022metric} to outperform the popular triplet-based losses. As can be seen in Table 2, our PMR loss shows stable performance boosts on both AudioCaps and Clotho datasets with different aggregation schemes. Besides, our PMR also achieves consistent performance gains on different baseline models, which further demonstrates the effectiveness of our PMR loss. 

\section{Conclusions}
\label{sec:conclusion}
In this work, we first highlight the drawbacks of text-agnostic audio pooling functions and then propose a text-aware attention pooling (TAP) module for text-audio retrieval. Our TAP can learn to attend to the most relevant frames to a given text while suppressing frames not described in the text, thereby enabling the model to flexibly extract the most semantically relevant information of the audio frames. Furthermore, we introduce a prior matrix revised (PMR) loss to revise the similarity matrix between the text and audio. By introducing a prior probability matrix calculated in the cross direction, the hard case with only single-side match can be filtered, thus producing more convincing dual optimal results. Experiments show that our TAP performs better than various text-agnostic pooling functions. Moreover, our PMR loss also shows stable performance gains on publicly available AudioCaps and Clotho datasets.

\bibliographystyle{IEEE.bst}
\bibliography{refs.bib}

\end{document}